\documentclass[preprint,showpacs,endfloats,pof]{revtex4}
\usepackage{amsfonts}
\usepackage{amsmath}
\usepackage{amssymb}
\usepackage{graphicx}
\usepackage{color}

\date{today}
\newcommand{\bq}{\begin{equation}}
\newcommand{\eq}{\end{equation}}
\newcommand{\ba}{\begin{array}}
\newcommand{\ea}{\end{array}}
\newcommand{\bqs}{\begin{equation*}}
\newcommand{\eqs}{\end{equation*}}
\newcommand{\bqa}{\begin{eqnarray}}
\newcommand{\eqa}{\end{eqnarray}}
\newcommand{\bqas}{\begin{eqnarray*}}
\newcommand{\eqas}{\end{eqnarray*}}
\newcommand{\bms}{\begin{bmatrix}}
\newcommand{\ems}{\end{bmatrix}}
\newcommand{\Rey}{\ensuremath{\mathrm{Re}}}
\newcommand{\Pe}{\ensuremath{\mathrm{Pe}}}

\newcommand{\Sc}{\ensuremath{\mathrm{Sc}}}

\newcommand{\pt}{\partial}

\begin{document}
\preprint{}

%\title[Chemically induced shear flow instabilities]{ImprovedDisplacement Efficiency via Chemically Induced Shear Instabilities}
\title{A novel shear flow instability triggered by a chemical reaction in the absence of inertia}
%\title{Observations of a Shear Flow Instability at Low Re Triggered by a Chemical reaction}\
\author{Teodor  Burghelea }\thanks{Corresponding author}
\affiliation{Department of Mathematics, University of British Columbia, 1984 Mathematics Road, Vancouver, BC, V6T 1Z4, Canada}

\author{Kerstin Wielage-Burchard}
\affiliation{Department of Mathematics, University of British
Columbia, 1984 Mathematics Road, Vancouver, BC, V6T 1Z4, Canada}

\author{Ian Frigaard}
\altaffiliation{Department of Mechanical Engineering, University
of British Columbia, 2054-6250 Applied Science Lane, Vancouver,
BC, V6T 1Z4, Canada} \affiliation{Department of Mathematics,
University of British Columbia, 1984 Mathematics Road, Vancouver,
BC, V6T 1Z4, Canada}

\author{D. Mark Martinez}
\affiliation{Department of Chemical and Biological Engineering, University of British Columbia, 2216 Main Mall, Vancouver, BC, V6T 1Z4, Canada.}

\author{James J. Feng}
\altaffiliation{Department of Chemical and Biological Engineering,
University of British Columbia, 2216 Main Mall, Vancouver, BC, V6T
1Z4, Canada.} \affiliation{Department of Mathematics, University
of British Columbia, 1984 Mathematics Road, Vancouver, BC, V6T
1Z4, Canada}

\keywords{Chemically Reactive Flows, Displacement Flows, Shear Flow Instability}
\pacs{47.20.-k Flow instabilities,47.50.-d Non-Newtonian fluid
flows ,47.57.Ng Polymers and polymer solutions,47.70.Fw Chemically
reactive flows}

\begin{abstract}
We present an experimental investigation of a novel low Reynolds
number shear flow instability triggered by a chemical reaction. An
acid-base reaction taking place at the interface between a
Newtonian fluid and Carbopol-$940$ solution leads to a strong
viscosity stratification, which locally destabilizes the flow. Our
experimental observations are made in the context of a miscible
displacement flow, for which the flow instability promotes local
mixing and subsequently improves the displacement efficiency.  The
experimental study is complemented by a simplified normal mode
analysis to shed light on the origin of the
instability.\end{abstract}

%\volumeyear{year}
%\volumenumber{number}
%\issuenumber{number}
%\eid{identifier}
%\date[\textbf{DATED: }]{\today}
%\received[Received text]{date}
%
%\revised[Revised text]{date}
%
%\accepted[Accepted text]{date}
%
%\published[Published text]{date}

\maketitle
%\tableofcontents{}

\section{Introduction}
\label{sec:intro}

Designing a method to locally control the hydrodynamic stability
of a shear flow is of importance in many practical and laboratory
applications. In the absence of inertia, non-equilibrium
hydrodynamical systems may lose stability if a relevant physical
field becomes stratified. To help illustrate this point,
convective motion may be induced in a thin static layer of fluid
when heated from below \cite{convection} or gravity induced
density stratification may sustain internal gravity waves,
\cite{Landau}. An alternative method of locally destabilizing low
Reynolds number ($\Rey$) shear flow would be to induce significant
changes in the local fluid rheology. Although flows with a
strongly stratified viscosity have been proved to be theoretically
unstable, \cite{Ern03} this situation has never been demonstrated
experimentally.  With flows of simple Newtonian fluids, it is
difficult to  vary the viscosity locally to induce an instability.
With complex (or structured) fluids ,however, the situation is
significantly different: the rheology is strongly coupled to the
molecular scale organization of the fluid. This opens a new
possibility of locally controlling the viscosity by inducing local
changes in the molecular structure via a chemical reaction. The
advantage of such method is that a chemical reaction may be
controlled by either mass transfer or by local heating or cooling.
In this paper we show experimentally that a displacement flow of
two miscible liquids may be destabilized by local changes in the
fluid rheology triggered by an acid-base reaction at their
interface.

Miscible displacements have been studied in depth within the
context of Hele-Shaw and porous media displacement instabilities,
initially motivated by a desire to better understand oil reservoir
recovery issues,
e.g.~\cite{Tan86,Hick86,Chan86,Yort87,Tan88,Yort88}, and this
science is now well developed. In the Navier-Stokes setting,
miscible displacements through small ducts have been studied both
experimentally and numerically,
\cite{Peti96,Chen96,Rako97,Yang97}. For the large P\'{e}clet
number ($\Pe$) regime, depending on the viscosity ratio~$m$,
quasi-steady viscous fingers may form and propagate with a
\emph{sharp} displacement front that is retained over long
timescales.
 The efficiency of the displacement (defined as the amount of fluid removed from the pipe walls)
 may be either measured or estimated using computation or asymptotic methods, see e.g.~\cite{Yang97,Laje99a}.
 In the case that the fingers remain stable, the residual wall layers thin
 at a rate that is significantly slower than the mean flow and the efficiency of the displacement is~$<1$.
 Although asymptotically the efficiency may approach~$1$ as~$t \to \infty$, for various practical reasons one might not want to wait.

To improve the displacement efficiency, one method would be to trigger
a flow instability, such that the perturbed interface disturbs the residual wall layers.
At large $\Rey$, multi-layer flows of viscous fluids are usually unstable, but linear interfacial
instabilities are also found for quite low $\Rey$. Dating from the late 1960's,
there are a number of studies involving both immiscible and miscible fluids, e.g.~\cite{Yih67,Hick71}.
 An extensive review can be found in the text \cite{Jose93}, and the physical mechanisms governing
 short and long wavelength instabilities have been explained by \cite{Hinc84,Char00}.
 These studies generally refer to the situation where there is a jump in the viscosity at the interface between two
fluids. If the change in viscosity is instead gradual, e.g.~due to a
diffuse interfacial region, then the stability characteristics
 appear to mimic those of the system with the viscosity jump, see \cite{Ern03}.

In porous media, non-monotone viscosity variations are known to cause
 linear instability of planar displacements, see \cite{Hick86,Chik88,Mani93},
 as can be predicted by classical mobility ratio arguments.
 Here however, we are in the Navier-Stokes regime and consider primarily shear flows. Also we have sharp localized
 change in viscosity, which is hard to achieve in ``simple'' fluids,
 where viscosity is often related to slowly varying concentration or temperature,
 i.e. due to molecular structure of the fluid.
 For complex (or structured) fluids the viscosity depends strongly on the microscopic structure.
 In the case of polymer solutions, the microscopic structure of the fluid can be locally modified by either
 mechanical means (e.g.~shear-thinning, \cite{Bird}) or chemical means,
 by locally modifying the chemical bonds between neighboring polymer molecules.

In recent years there have been a number of studies of systems
with coupled chemical reactions and fluid flow, so it is natural
to examine the relation to this literature. In the first place
even without fluid flow, spatially traveling waves can be
observed in chemically excitable media governed by coupled
reaction-diffusion systems, see e.g.~the review \cite{Tyso88}.
These fronts may frequently destabilize linearly. There is often
coupling between the reactions, and hence a mechanism of feedback,
and sometimes a significant difference in the reaction rate
constants. Slightly simpler systems involve single species
auto-catalytic reactions, which mathematically admit traveling (chemical)
wave solutions. Systems in which the auto-catalytic reaction
results in a significant density change have been studied
intensively. The base traveling chemical wave is coupled with a
Rayleigh-Taylor problem. There are consequently a wide range of
stable and unstable situations to be studied, and even more once a
base fluid velocity is considered. A selection of the many
works includes
\cite{Pojman90,Edwards91,Vasquez91,Vasquez94,Edwards02,Vasquez04,Hern06,Lima06}.
Of closer relation to our work is the sequence of papers,
\cite{DeWit97a,DeWit97b,DeWit99a,DeWit99b}, which have considered
a concentration dependent viscosity in the context of miscible
porous media displacements. Although interesting and a useful
guide in methodology, direct relevance of these many papers to our
study is in fact limited. Our system is not auto-catalytic, there
are no buoyancy effects and the displacement flow is not a
gradient flow.

\subsection{Industrial motivation}
\label{sec:industrial}

The motivation for our study comes from the construction of oil and gas wells.
Since the early 1990's there has been an increasing number of wells that are constructed with long horizontal sections.
The worlds longest extended reach wells have horizontal sections in the $10-15$~km range, but these are exceptional.
 More routinely, wells are built with horizontal extensions of up to $\sim 7$~km.
One of the key barriers in constructing longer wells comes from
simple hydraulic friction. In a vertical well, both the pore
pressure of reservoir fluids and the fracture pressure of the
reservoir rock increase with depth, approximately linearly.
Judicious choice of fluid density and circulating flow rates keeps
the wellbore pressure inside the so-called ``pore-frac envelope'',
i.e. the region where the porous rock does not fracture. In a
horizontal well section, the pore-frac envelope is unchanged with
length along the well, but the frictional pressure increases with
length, leading to eventual breaching of the envelope
\cite{Bourg,Porefrac,Horiz}.

There is a consequent interest in methods and fluids that control
the frictional pressure in some way. Two operations where this is
important are drilling and cementing. In the cementing process,
\cite{Nels90}, it is necessary to displace the drilling mud with a
spacer fluid and then with a cement slurry. As the section is
horizontal, density differences between the fluids lead to
stratification and should be avoided. Instead the focus is on
controlling the rheology of the fluids and the displacement flow
itself.
 The idea behind the reactive instability that we are studying is explained by the following simple calculation.

Suppose simplistically that we have Newtonian fluids, a circular pipe of radius $\hat{R}$
and that we wish to displace at mean speed $\hat{U}_0$. If $\Delta \hat{p}_e$ is the difference between pore and fracture limits,
then we are restricted to a length-viscosity combination:
\[ 8 \frac{\hat{L} \hat{\eta} \hat{U}_0}{\hat{R}^2 } < \Delta \hat{p}_e. \]
Normally, we will need to select the viscosity of the spacer fluid, $\hat{\eta}_s > m \hat{\eta}_d$, where $m > 1$ and $\hat{\eta}_d$
is the viscosity of the in-situ fluid, e.g.~drilling mud, so that
\begin{equation}
\hat{L}  < \frac{ \hat{R}^2 \Delta \hat{p}_e }{8 m \hat{\eta}_d \hat{U}_0 }.
\end{equation}
Suppose instead we are able to pump a spacer fluid of viscosity
$\hat{\eta}_s \leq \hat{\eta}_d$, that reacts with the in-situ
fluid to cause a local instability. If the instability causes
effective mixing across the pipe after the fluid has traveled $k$
radii, we may model this process by a radial diffusivity
$\hat{D}_r \sim (\hat{R}\hat{U}_0)/k$. Provided $k \ll
\hat{L}/\hat{R}$, we have an effective Taylor dispersion process
in which the mixed zone diffuses axially along the pipe, relative
to the mean flow. This axial diffusion of the mixed region is
governed by a dispersion coefficient of order $\sim k D_T
(\hat{R}\hat{U}_0)$, where $D_T = 1/48$ is the Taylor dispersion
coefficient for a pipe. After traveling a distance $\hat{L}$, the
mixed zone will have dispersed axially a distance
\[ \hat{L}_{\mathrm{mix}} \sim \left[ k D_T \hat{R}\hat{L} \right]^{1/2} . \]
If the reacted mixture has viscosity $(a+1) \hat{\eta}_s$, where $a\ge 0$,
then the frictional pressure drop along the pipe can be limited by
\[ \Delta \hat{p}_e > 8 \frac{\hat{\eta}_s \hat{U}_0}{\hat{R}^2} \left(  \hat L + a\hat L_{\mathrm{mix}}\right) , \]
and the pipe length restriction is
\begin{equation}
\hat{L} < \frac{\hat{R}^2 \Delta \hat{p}_e }
{ 8 \hat{\eta}_s \hat{U}_0 ( 1  +  a [k D_T \hat{R}/ \hat{L}]^{1/2} )} .
\end{equation}
Therefore, for a given pore-frac limit, we may increase the length of the well that may be effectively displaced provided that:
\begin{equation}
\frac{m \hat{\eta}_d }
{\hat{\eta}_s ( 1  +  a [k D_T \hat{R}/ \hat{L}]^{1/2} )}  > 1.
\end{equation}
Over lengths sufficiently long that $a [k D_T \hat{R}/ \hat{L}]^{1/2} \ll 1$, we achieve a length
increase by a modest factor $m \hat{\eta}_d/ \hat{\eta}_s $.

Even modest increases in length may have a large impact, both economically and environmentally.
In modern offshore drilling, one vertical well drilled down from the seabed may act as the stem
for many lateral horizontal branches, extending radially outwards.
 Thus, increases in the length of horizontal branches correspond to (length)$^2$
 increases in the area of reservoir that may be reached and contribute to a reduction
  in the number of wellheads required. This reduces the cost of the field development reduces
  the environmental footprint, lessens risks of leakage of reservoir fluids and eventually makes for an easier well abandonment.

\subsection{Organization}
\label{sec:organ}

Our paper is organized as follows. Section \ref{sec:expdesc}
describes the experimental setup, the measurement techniques and
the rheology of our fluids. Experimental results are presented in
\S\ref{sec:expresults}. Section \ref{sec:toymodel} introduces a
simple hydrodynamic stability model that gives insight into the
instabilities we observe. The paper closes with a brief discussion
of our findings and a discussion on future theoretical and
experimental studies.

\section{Description of the experiments}
\label{sec:expdesc}

\subsection{Experimental apparatus and techniques}

\begin{figure}[h]
\includegraphics[width=0.95\textwidth]{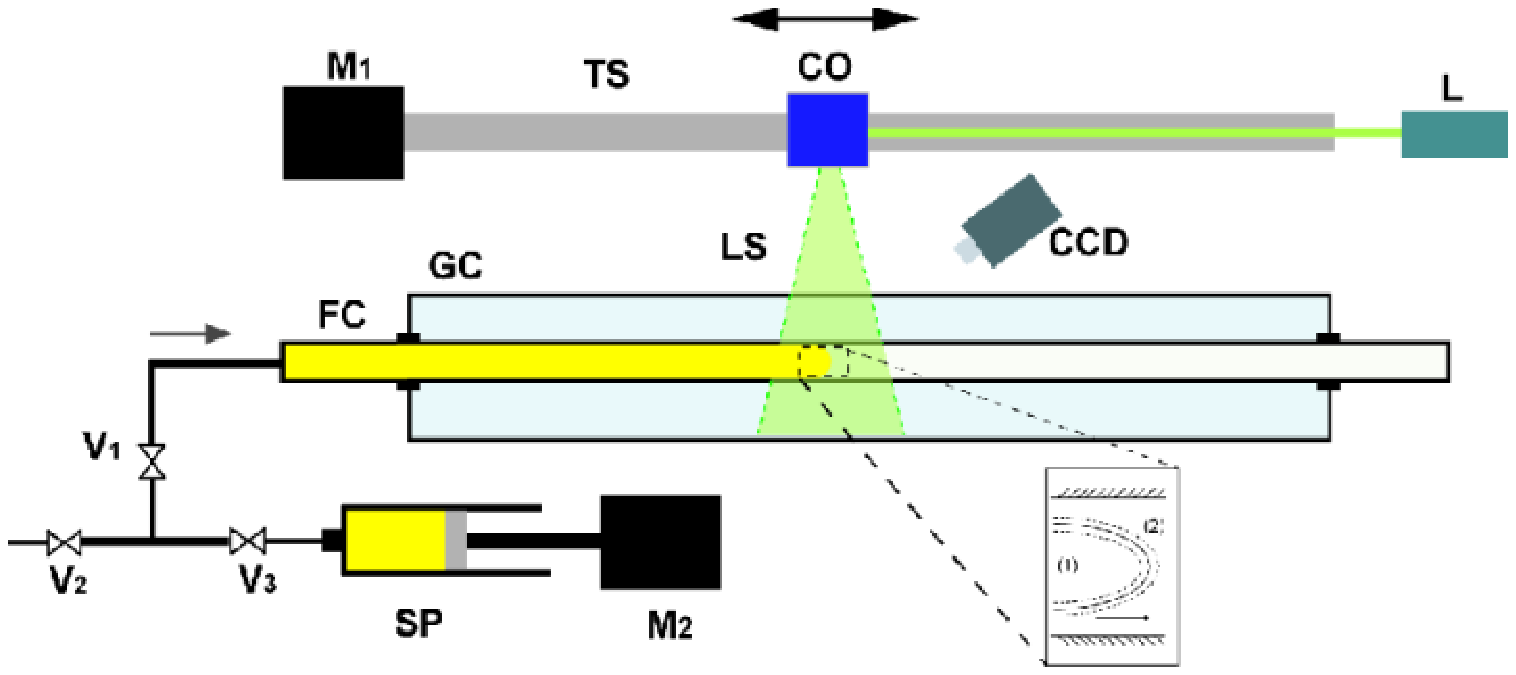}
\caption{Schematic view of the experimental apparatus:
\textbf{FC}-flow channel, \textbf{GC}-glass container,
\textbf{L}-solid state laser, \textbf{CO}-cylindrical optics,
\textbf{LS}-laser sheet, \textbf{CCD} -camera,
$\mathbf{M_{1,2}}$-stepping motors, $\mathbf{V_{1,2,3}}$-valves,
\textbf{TS}-horizontal translational stage,  \textbf{SP}-syringe
pump. The lower inset illustrates schematically the flow
configuration: the full curve represents the interface between
fluids, the dotted lines indicate the reacting region and the
horizontal arrows point the mean flow direction.} \label{f.1}
\end{figure}

All experiments were conducted in the apparatus illustrated
schematically in Fig.~\ref{f.1}. It consists of a horizontal flow
channel \textbf{FC} with  circular cross section of radius $\hat
R=5.5$~mm and length $\hat L=1.2$~m, immersed in a water filled
glass container \textbf{GC} to ensure distortion-free flow
illumination and imaging. The system was illuminated by a thin
laser sheet passing horizontally through the transparent walls of
both the glass container and the flow channel, at the middle
vertical position. The laser sheet has a thickness of
approximately $50~\mu$m in the center of the set-up and about
 $160~\mu$m near the walls of the channel. It was generated by passing a laser beam delivered by a $150$~mW solid state laser,
 \textbf{L}, through a block of two crossed cylindrical lenses, \textbf{CO}, mounted in a telescopic arrangement.

The flow was imaged from the top (Fig.~\ref{f.1}) with a charge
coupled device camera, \textbf{CCD},
 equipped with a $35$~mm photographic lens.
 The images were digitized with $8$ bit quantization and $640\times 760$ pixels resolution (which accounts for $130~\mu$m).
The size of the imaged area was thus $1.4\times 1.4~\mathrm{cm}^2$.

The flow in the channel was induced by a syringe pump \textbf{SP}
actuated by a precise stepping motor, $\mathbf{M_1}$ (from $CBVL$,
Vancouver) and controlled by computer via a serial port. The inflow
mean velocity was controlled with an accuracy better than
$100~\mu$m/s. In order to monitor the evolution of the fluid
interface downstream, the \textbf{CCD} and the cylindrical optics
block, \textbf{CO}, were mounted on a linear translational stage
parallel to the flow channel (\textbf{TS} in Fig.~\ref{f.1}). The
stage was actuated by a stepping motor $\mathbf{M_1}$ and controlled
by computer via a serial port. The typical configuration of fluids
during the horizontal displacement experiments is schematically
illustrated in the inset of Fig.~\ref{f.1}.

The stability of the interface between fluids was investigated
using either Laser Induced Fluorescence ($LIF$), or Digital
Particle Image Velocimetry ($DPIV$).  The image acquisition
software was developed in-house and allowed us to adjust the time
delay between successive frames in relation to the local flow
velocity (in order to keep the mean particle displacement in the
range $5-30$ pixels). For low values of the flow velocity the time
delay was $66$~ms. For higher flow speeds, the delay was decreased
to $16$~ms. The fluids were seeded with either a small amount
(approx. $160$~ppm) of $20~\mu$m Polyamide spheres (from Dantec
Inc.) for $DPIV$ measurements, or with fluorescein sodium salt
(Sigma Aldrich) for the $LIF$ measurements. Time series of the
velocity fields were obtained by a multi-pass $DPIV$ algorithm
\cite{Scarano}. The spatial resolution was $167~\mu$m. The
accuracy of the method has been carefully checked by running test
measurements for low Reynolds number Poiseuille flows with
Newtonian fluids and comparing then with the analytical solution.

In addition to these techniques, we have developed a third
visualization method to measure the local pH in the flow field. We
do so by using a $pH$ sensitive colored dye, Bromothymol Blue (Sigma
Aldrich). The local value of the $pH$ near the interface was
assessed by measurements of the color distribution in the field of
view. With this we are able to estimate the local rheological properties.

\subsection{Fluid properties}

We have used the same base fluids for all our experiments, with
various adjustments to the levels of acid and base used in each
fluid to examine different regimes. The displacing fluid,
Fluid~$1$, was a $65 \%$ aqueous sucrose solution. The pH of this
fluid was varied between~$7$ and~$11$ by titration with different
amounts of $NaOH$ ranging from $150$ parts per million ($ppm$) and
$~300 ~ppm$. The displaced fluid, Fluid~$2$, was a mixture of a
5\% (wt) Carbopol~940~(\text{C-940}) solution and a $65\%$ (wt)
sucrose solution in deionized water. C-940 is generally referred
to as a weak polyacrylic acid and dissociates in solution. The
$pH$ of this solution is typically between two and three. During
our experiments, neutralization of C-940 molecules present in
Fluid $2$ occurs locally, in the vicinity of the interface,through
contact with Fluid $1$.

\begin{figure}[h]
\includegraphics[width=0.55\textwidth]{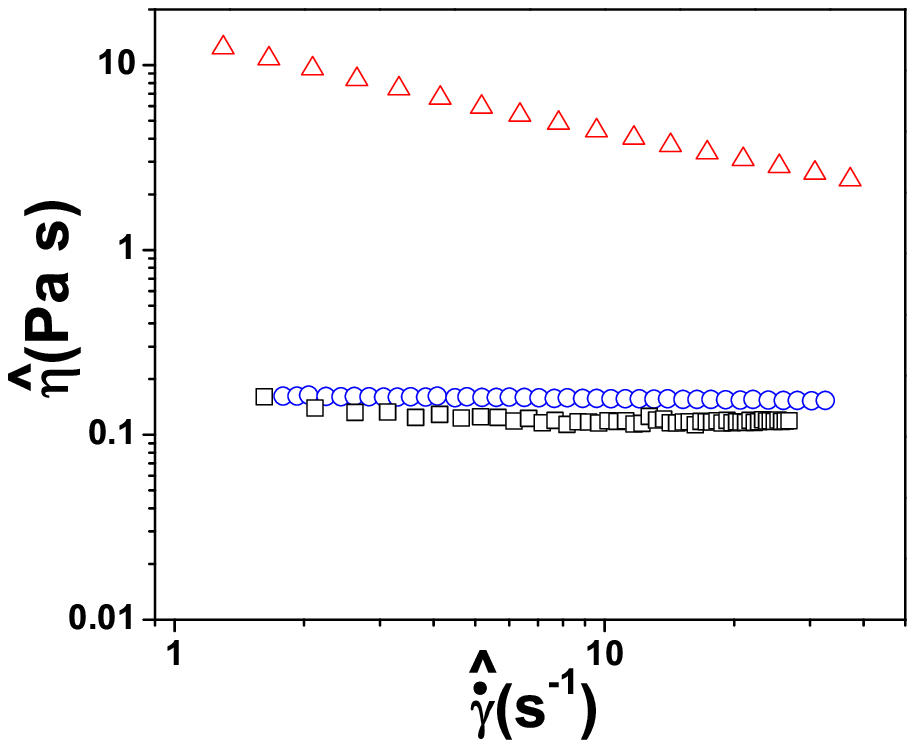}
\caption{Shear rate dependence of the viscosities of the fluids:
({\textcolor{blue}{$\circ$}})~displaced fluid~2 with $pH \approx
3$, ({\textcolor{black}{$\Box$}})~displacing Fluid 1 ,
({\textcolor{red}{$\triangle$}})~displaced Fluid 2 with $pH
\approx 7$. } \label{f.2}
\end{figure}

The $pH$ of the fluids were measured with a thermally corrected
digital $pH$ meter (Topac Instruments) with $1.5 \%$ accuracy. The
rheological properties of both fluids~$1$ and~$2$ were measured
with a stress controlled rotational rheometer: $CVOR$ acquired
from Bohlin Instruments (Malvern Inc., www.malvern.co.uk). Example
flow curves at $22^{\circ}C$ of Fluid~$1$, Fluid~$2$ (at $pH
\approx 3$) and Fluid~$2$ (at $pH \approx 7$), are presented in
Fig.~\ref{f.2}. As shown, the viscosity of Fluid~$1$ is
independent of shear rate. With Fluid $2$ at a $pH \approx 3.2$,
the viscosity is also approximately constant. As a guide, at a
shear rate of 2 $s^{-1}$ we have viscosities
$\hat\eta_{1}=114$~mPas and $\hat\eta_{2}=138$~mPas, for fluids
$1$ and $2$ respectively. As the $pH$ is increased the rheology of
Fluid $2$ changes dramatically (see Fig.~\ref{f.2}). A detailed
study of the coupling between the $pH$, the molecular structure
and the rheological properties of C-940 has been recently reported
in \cite{Frisken}.

\begin{figure}[h]
\includegraphics[width=0.85\textwidth]{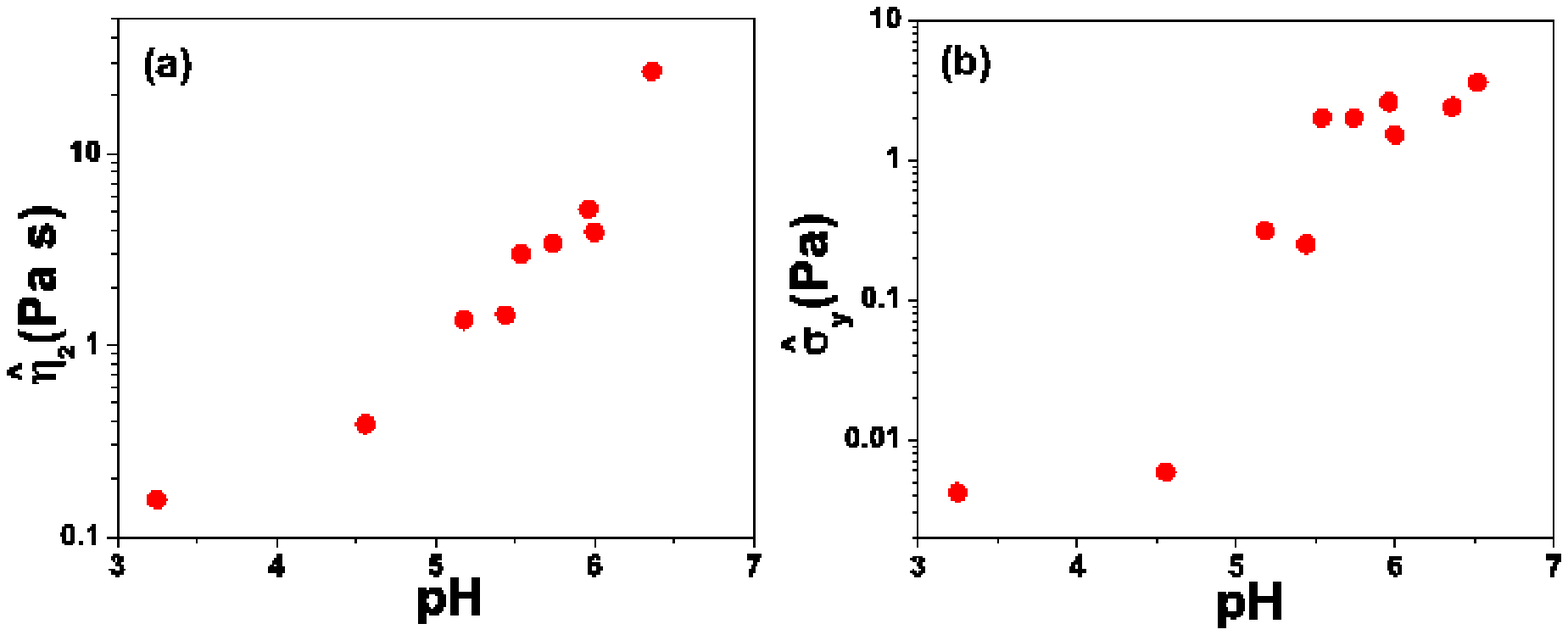}
\caption{Dependence of the rheological properties of fluid $2$ on
$pH$: (a) viscosity measured at $\dot{\gamma}=2s^{-1}$ (b) yield
stress. } \label{f.3}
\end{figure}

In order to quantify the dependence of the rheological properties
of Fluid $2$ on the $pH$, several batches have been prepared at
different $pH$ values, and the rheological tests have been
conducted. The $pH$ dependence of the shear viscosity of fluid $2$
at rate of strain $\dot{\gamma}=2~s^{-1}$, is shown in
Fig.~\ref{f.3}a. In the range of $pH$ shown, the shear viscosity
increases monotonically, up to a value nearly two orders of
magnitude larger than the non-neutralized value. The $pH$
dependence of the measured yield stress of Fluid $2$ is shown in
Fig.~\ref{f.3}b. A further increase of the $pH$ (data not shown in
Fig.~\ref{f.3}) results in destruction of the gel structure, which
causes both viscosity and yield stress to drop to their initial
low $pH$ values.

\subsection{Experimental procedure}

Our displacement experiments were conducted as follows. First, the
alignment of the laser sheet and focus of the camera is carefully
checked at several locations downstream. Second, the valves $V_1$,
$V_2$ are open and  the channel is initially filled with Fluid
$2$. Next, we close valve $V_1$, open valve $V_3$ and start to
drive the syringe pump at very low speeds for several tens of
seconds. This procedure allows us to eliminate any air bubbles
generated during the filling of the flow channel. Finally, we
close valve $V_2$ and open valves $V_1$ , and set the fluids in
motion by operating the syringe pump at the desired speed. The
flow images were usually acquired halfway down length of the pipe,
though in some of the experiments the camera and the laser sheet
were moved at constant speed along the flow channel. Approximately
$1000$ flow images were acquired both before the entrance of the
fluid interface in the measuring window and after. Details of the
experimental conditions tested are given in
Table~\ref{experiments}.

\section{Experimental results}
\label{sec:expresults}

We propose to control the stability of an interface between a
Newtonian and a C-940 solution by inducing changes in the local
fluid rheology via an  acid-base reaction, which depends on the
free charge mismatch across the interface,and the $pH$ of each
fluid. With $1\%$~NaOH, the saccharose solution (Fluid~1) has a pH
around~$11$. When in contact with Fluid $2$ at $pH \approx 3$,
neutralization occurs near the interface resulting in a thin layer
of high viscosity fluid, see Fig. \ref{f.2}. The high viscosity
interfacial layer apparently undergoes a self-sustained
hydrodynamic instability that results in local mixing of the two
fluids.
 Below we will present a
full description of this reactive instability and displacement.
For comparison, we have conducted~$2$ experimental test sequences
(at varying flow rates) in which there is no reaction, see
Table~\ref{experiments} series~1 and~2. In addition we have
performed a number of reactive experiments, see series $3$ in
Table~\ref{experiments}. In total, we performed $5$ experiments in
this series.

\begin{table}[h]
\centering
\begin{tabular*}{\textwidth}{@{\extracolsep{\fill}}|l|c|c|*{6}{c|}}
\hline
            & \multicolumn{2}{c|}{\textbf{Fluid 1}}
            & \multicolumn{4}{c|}{\textbf{Fluid 2}}
            & &\\
            & \multicolumn{2}{c|}{\textbf{Composition:} }
            & \multicolumn{4}{c|}{\textbf{Composition:} }
            & & \\
            & \multicolumn{2}{c|}{$65~\%$ saccharose }
            & \multicolumn{4}{c|}{$66~\%$ saccharose}
            & & \\
    \textbf{Exp.}
            & \multicolumn{2}{c|}{$\mathbf{\hat\eta_1} = 114 $ mPas}
            & \multicolumn{4}{c|}{~}
            & & \\
    \textbf{Sequence}
            & $\mathbf{pH}$&\textbf{Re}
            & \textbf{C-940~(\%)}
            & $\mathbf{pH}$
            & \textbf{Re}
            & \textbf{$\mathbf{\hat\eta_2}$~(Pas)}
            & \textbf{$\mathbf{\hat Q}$~(ml/s)}&\textbf{Stab.}\\
\hline \hline
 \textbf{Control} $\mathbf{1}$ & $7$& $0.03-0.18$& 0      &$7$& $0.03-0.15$          & $0.138$&$0.06-0.3$& \textbf{S}\\
 \hline
 \textbf{Control} $\mathbf{2}$ & $7$& $0.02-0.3$ & $0.1$ &$7$& $\left(0.1-1.5\right)\cdot 10^{-03}$& $20$   &$0.03-0.4$& \textbf{S}\\
\hline
\textbf{Reactive} & $7.7-11.5$      &$0.07-0.2$  & $0.1$ &$3$& $0.05-0.3$        & $0.144$&$0.1-0.5$& \textbf{U}\\
\hline
\end{tabular*}
 \caption{The experiments and the corresponding fluids
properties. For each fluid, $\Rey$ was calculated using the $DPIV$
measured mean flow velocity, the corresponding viscosity
coefficients and the densities $\rho_1 \simeq \rho_2\sim
1285$~kg/m$^3$. The flow rate $\hat Q$ has been estimated via the
$DPIV$ measured mean flow velocity. In the last column of the
table, the symbols \textbf {S}, \textbf{U} stand for stable and
unstable, respectively.} \label{experiments}
\end{table}

\subsection{Control experiments}

\begin{figure}[h]
\includegraphics[width=\textwidth]{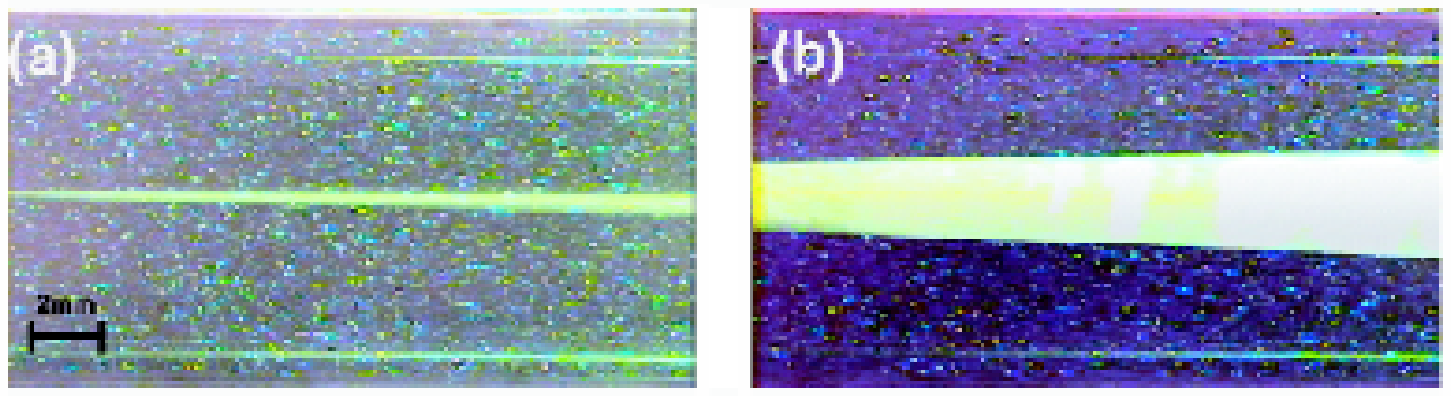}
\caption{Example fluorescent images of the interface in an
experiment from control sequence~1: (a-b) Fluid $1$~-~$65\%$
saccharose solution, Fluid $2$~-~$66\%$ saccharose solution, flow
rate $\hat{Q} =0.145$~ml/s. The two images are separated in time
by $5$~s.} \label{f.4}
\end{figure}

The first control sequence concerns the two base fluids without
reaction. We have seen in Fig.~\ref{f.2} that the viscosity of
Fluid 2 at $pH \approx 3$ is largely constant and governed by that
of the underlying saccharose solution. We therefore conduct a
control displacement of Fluid~$1$, a $65\%$~saccharose solution,
by Fluid~$2$, a $66\%$ saccharose solution. The flow rates are
chosen so that $\Rey < 1$ in all cases, and often $\Rey \ll 1$.
Both fluids are at neutral $pH$. The displacing fluid is mildly
more viscous, but not sufficiently so to ensure a good
displacement, ($\hat\eta_{1}=114$~mPas and
$\hat\eta_{2}=138$~mPas) as illustrated in Figure~\ref{f.4}. The
interface develops into a long finger that stretches progressively
along the pipe. At no time do we observe any interfacial
instability, and the interface remains sharp, i.e.~the length of
the tube is insufficient for molecular diffusion to have any
significant effect. Qualitatively similar results for miscible
Newtonian displacements can be found in \cite{Peti96}.

\begin{figure}[h]
\includegraphics[width=\textwidth]{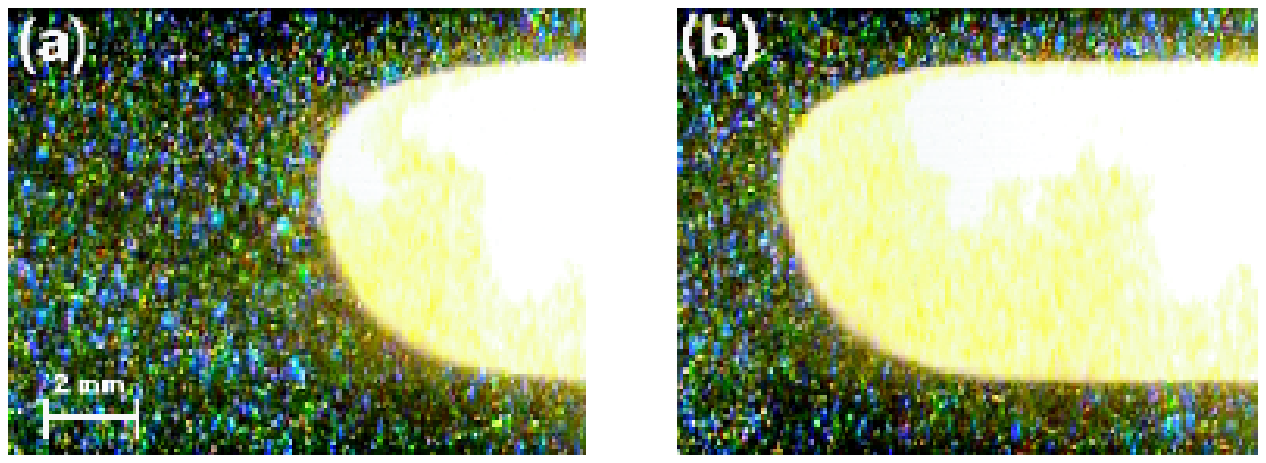}
\caption{Example images of the interface in an experiment from
control sequence $2$ ($\hat Q=0.08$~ml/s). Fluid $1$ has been
doped with fluorescein for visualization of the fluids interface.
} \label{f.5}
\end{figure}

The second control sequence also concerns displacement of the two
fluids at neutral $pH$, i.e. without chemical reaction, but
Fluid~$2$ has a much larger viscosity ($\hat \eta_2 \approx
20$~Pas at $\dot \gamma=1~s^{-1}$) and a significant yield stress
($\hat \sigma_y \approx 6.6$~Pa). The viscosity of Fluid~$1$ is
unchanged. Similar flows have been studied experimentally before
in slightly larger tubes \cite{Gaba01,Gaba03}. As commented
in~\cite{Gaba01}, some care is needed in choosing the displacing
flow rate, since at low flow rates the displacing fluid
effectively fractures through the elastic gel, leaving a rough
edge. For low flow rates we also observed a slightly granular
interfacial texture and also some asymmetry of the finger. The
slight asymmetry of the finger visible in Fig.~\ref{f.5} could be
due to the non-symmetric entrance condition (the interface between
fluids enters the flow channel via a $T$ shaped junction, as shown
in Fig.~\ref{f.1}) which is preserved at all later times due to
the significant yield stress of Fluid~$2$. Alternatively the
asymmetry may relate to a transition between solid-like and
fluid-like behavior. We do not attempt to explain this further.

Sample images of an experiment in this sequence are shown in
Fig.~\ref{f.5}. The C-940 solution yields in the center of the
pipe, where there is a two-dimensional flow, but the stresses are
not sufficient to make it yield at the wall. Consequently a static
residual wall layer is left in the tube as a finger of Fluid~1
penetrates steadily along the pipe. The shape of the nose of the
finger is quite rounded and the wall layers have apparently
constant thickness.

Over the duration of the experiment, there is no evidence of any
interfacial instability. These results are qualitatively similar
to those of \cite{Gaba01,Gaba03}, where C-940 solutions are
displaced by glycerol solutions. The reason that there is no
interfacial instability here is because the residual layers of
C-940 solution are in fact fully static and unyielded. Such flows
have been studied in some detail, theoretically and
computationally \cite{Allo00,Frig01a,Frig01c,Frig02}, and are well
understood. Even with a non-zero flow rate of C-940 solution,
stable regimes may be predicted and found experimentally, see
\cite{Moye04a,Huen06}.

The two control sequences establish that without a chemical
reaction and accompanying local rheology change, these
displacement flows are stable. In other words, having a change in
bulk rheology of the two fluids at an interface can result in
instability but does not for the systems we study.

\subsection{Chemically reactive unstable flows}

\begin{figure}[h]
\includegraphics[width=\textwidth]{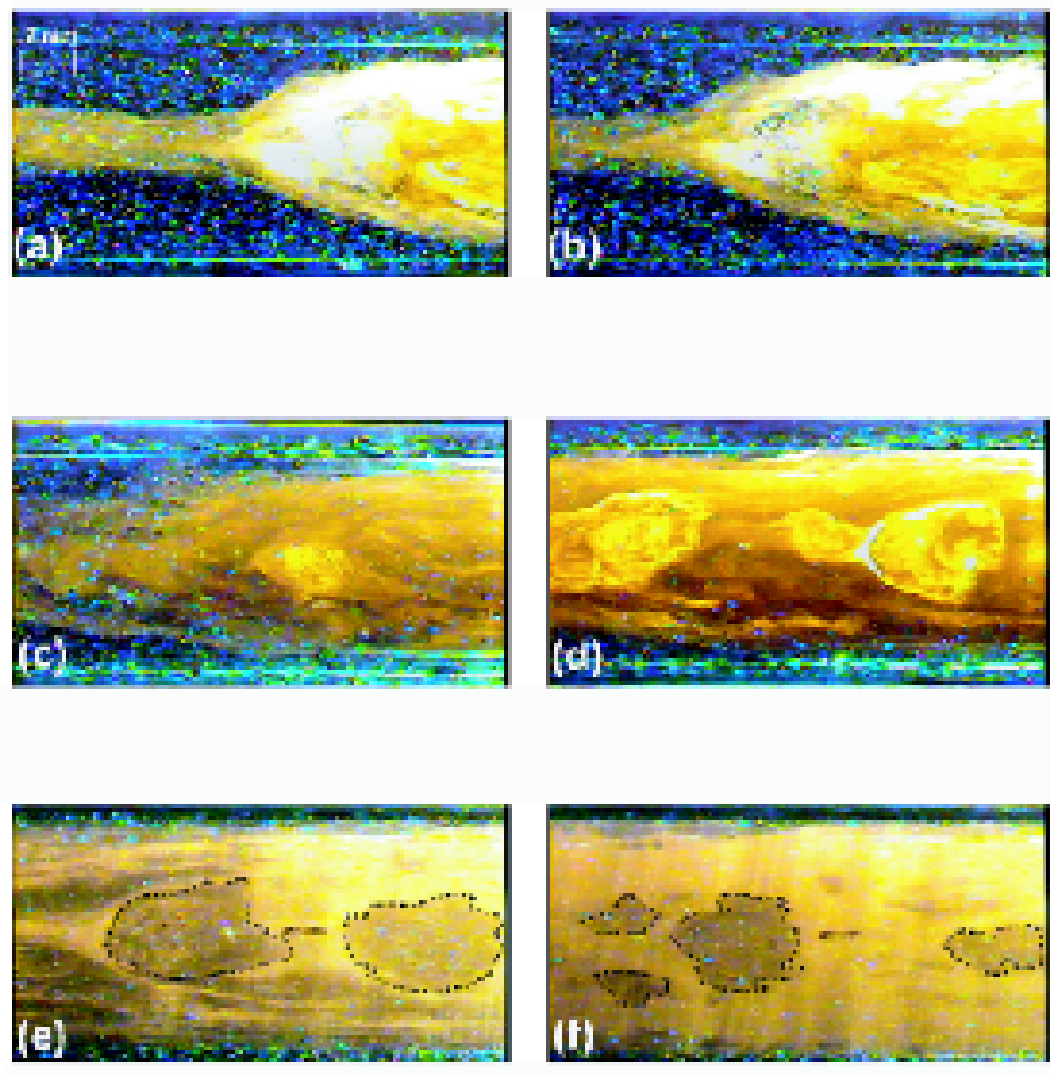}
\caption{(a-d)~Fluorescent images of the interface in a reactive
displacement: Fluid~$1$ - $65\%$~saccharose solution, Fluid~$2$ -
$0.1 \%$~C-940 in $66\%$ saccharose solution. (e-f)~Fluorescent
flow images long after the entrance of the unstable interface in
the field of view; the images are separated in time by
approximately $5$~s. The dotted lines highlight gelled structures
tumbling downstream.} \label{f.6}
\end{figure}

The flow behavior was significantly different from the control
experiments in the reactive case, when the C-940 solution at
$pH=3$ was displaced by a saccharose solution at $pH=11$, at
different flow rates. The initial interface penetrates in a sharp
spike as before, but this is destabilized and the finger rapidly
widens to nearly fill the pipe. A complex secondary flow develops
at the interface between fluids. The flow seems to be dominated by
large vortices advected by the flow, with a typical size of the
order of the pipe radius. Typical images are shown in
Fig.~\ref{f.6}.

\begin{figure}[h]
\includegraphics[width=0.95\textwidth,height=0.4\textheight]{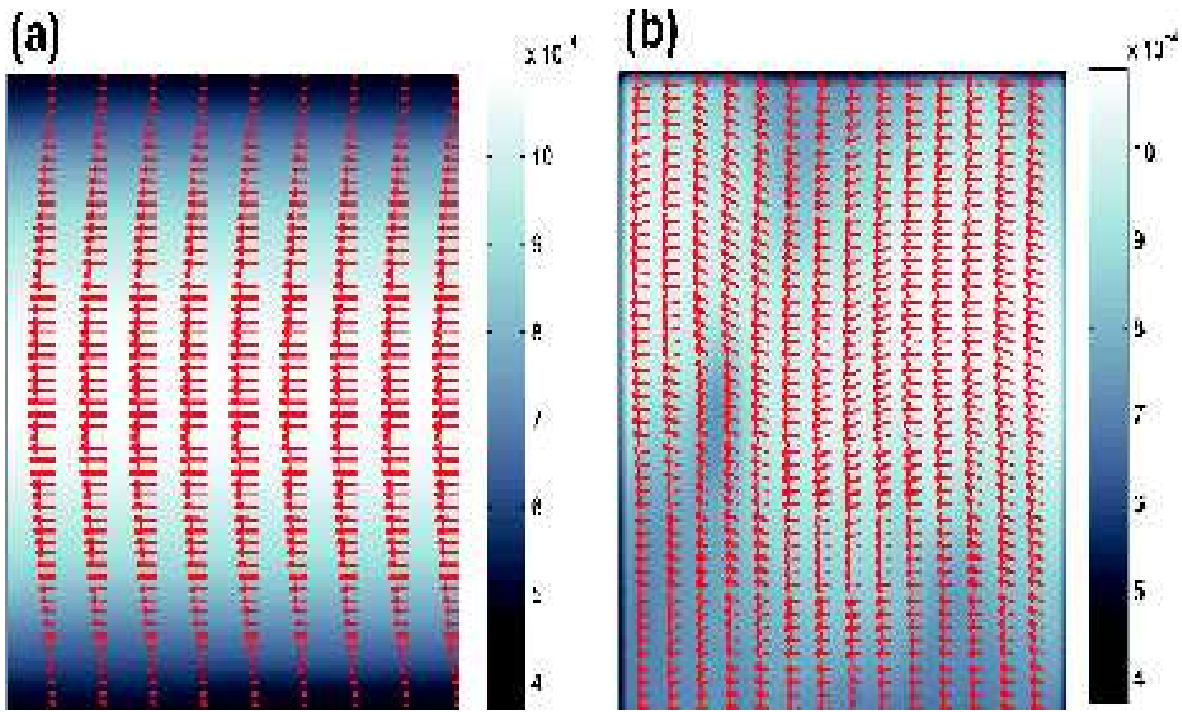}
\caption{Instantaneous velocity fields for a chemically reactive
displacement experiment, $\Rey \approx 0.15$:~(a) before entrance
of the finger in the field of view (b) after the entrance of the
finger in the field of view. The false color maps represent the
magnitude of the velocity. In order to enhance the clarity, we
only plot half of the vectors and re-scale the length of the
arrows. } \label{f.7}
\end{figure}

\begin{figure}[h]
\includegraphics[width=0.95\textwidth]{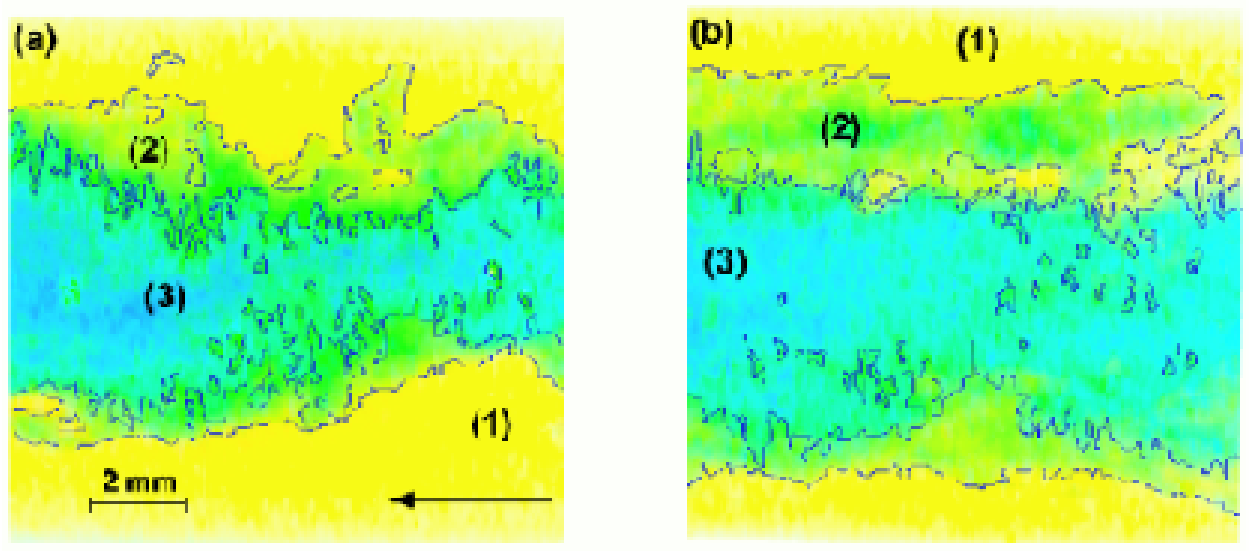}
\caption{Iso $pH$ regions near the interface: yellow~(region $1$)-$pH
\approx 3$, green~(region $2$)-$pH \approx 7$, blue~(region $3$)-$pH
\approx 11$. } \label{f.8}
\end{figure}

As the front of the finger passes, the secondary flow
instabilities persist along the sides of the finger. The secondary
flow provides a feedback mechanism for the instability by bringing
into contact new un-reacted fluid elements and taking away reacted
highly viscous fluid. The initial pass of the finger front does
not remove all the fluid~$2$ from the walls. However, the
secondary flows result in a fairly rapid erosion of the residual
layers. After the initial instability, small parcels of Fluid~2
pulled into the Fluid 1 stream react to form gelled solid regions
that are advected along with the fluid. Close observation of video
images reveals that some of these parcels appear to be in rigid
motion, see e.g.~Fig.~\ref{f.6}e \& f.

More detailed information on the structure of velocity field is
obtained from $DPIV$ images. We display in Fig.~\ref{f.7}a \& b
two instantaneous velocity fields measured during a chemically
reactive displacement experiment at roughly the midpoint of the
fluid channel downstream. Before the entrance of the displacement
front into the field of view, the flow is similar to
Hagen-Poiseuille flow, see Fig.~\ref{f.7}a. The velocity profile
is parabolic and there is no secondary fluid motion in a direction
orthogonal to the mean flow. Analysis of time series of velocity
fields (data not shown here) has shown that the time fluctuations
are only due to instrumental noise, which accounts for less than
$5 \%$ of the mean flow velocity. The structure of the flow field
changes drastically after the passage of the displacement front
through the field of view. As one can clearly see in
Fig.~\ref{f.7}b, the fluid motion follows a wavy-spiral pattern,
with an apparent periodicity in the axial direction of roughly the
tube diameter. The velocity field is now unstable and
characterized by a rather strong radial component. Quasi-periodic
entrance of slowly moving flow regions in the field of view can be
associated with the passage of fully gelled solid parcels of
fluid~$2$ illustrated in Fig.~\ref{f.6}e \&f and discussed above.

Finally, we are able to confirm that the mixing process is locally
effective and that the $pH$ varies over some intermediate range in
the displacement region. To do this we image unstable flows doped
with $pH$ sensitive colored dye, as shown below in Fig.~\ref{f.8}.
Since the $pH$-dependence of the fluid rheology has been
characterized, in principle we could use the $DPIV$ images
together with the data in Fig.~\ref{f.8} to construct local
deviatoric stress fields. This technique however still requires
some work.

\subsection{Displacement efficiency}

A more quantitative assessment of the displacement efficiency is
obtained by processing $LIF$ images to give an indication of the
evolution in time of the finger width at a fixed position along
channel, roughly $50$~cm from the entrance. In each experiment a
long time-series of images ($\sim 2000$) is acquired, starting
long before the entrance of the displacement front into the field
of view and ending long after its passage. In these experiments
fluid $2$ contained Polyamide spheres and fluid $1$ a small amount
of fluorescein (see section~\ref{sec:expdesc}~C). The images prior
to the entrance of the finger in the field of view are passed to
the $DPIV$ algorithm to obtain a time series of velocity fields.
The time average of these
 images also provides a background image that is used to compensate for non uniform
 illumination of each image in the sequence. The fluorescent images of the displacement
 front/finger are digitally processed to extract the width (measured along the radial direction)
 of the finger. Each image is first converted to a binary image, and the edge of the interface
  is detected using the $Sobel$ algorithm with a fixed brightness threshold. Spurious edge
  identifications, i.e.~objects of the order of several pixels that are identified due to
   brightness inhomogeneities, are carefully removed using a
morphological filter. Finally, the width of the finger is quantified from the width of the finger contour.

During reactive experiments, the instantaneous shapes of the
finger are usually non-symmetric in both azimuthal and radial
directions. However statistically, (considered as an ensemble
average over several positions of the finger), the symmetry does
not seem to break. Thus, the finger width measurement can be
interpreted as a volumetric measure of the displacement
efficiency, although we prefer to display time series of the
width. In Fig.~\ref{f.9} we display the time dependence of the
width of the finger for several values of flow rate. For different
flow rates the time until the displacement front enters the field
of view will be different. This residence time has been subtracted
and the time (on the horizontal axis) is then scaled with the
advection timescale, $\hat{R}/\hat{U}_0$. The finger width is
normalized by the diameter of the pipe, $\hat d=2\hat R$.

\definecolor{brown}{rgb}{0.545,0.27,0.185}
\begin{figure}[h,t]
\includegraphics[width=\textwidth]{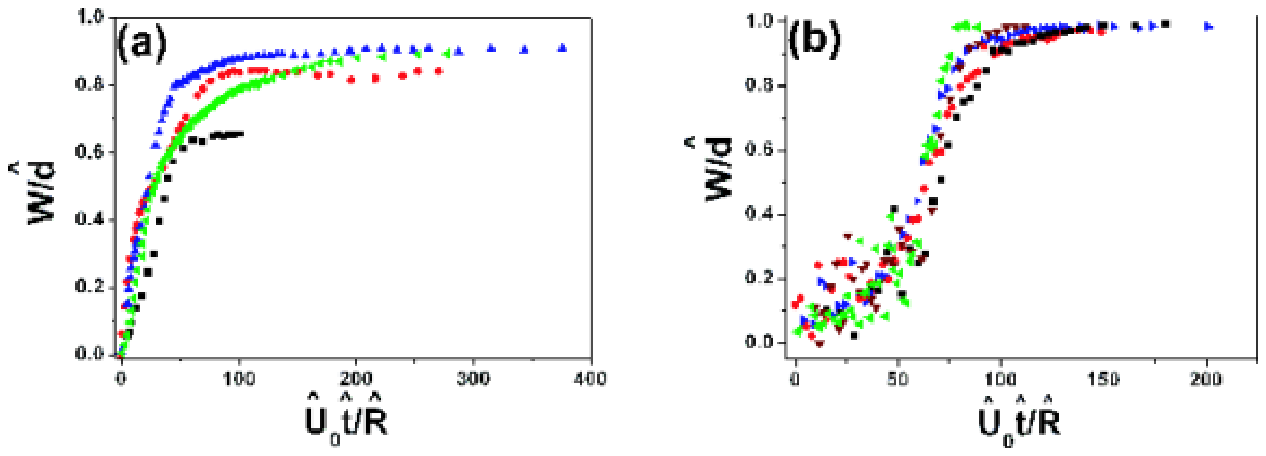}
\caption{(a) Normalized width of the tip versus the normalized
displacement distance, $\hat U_0 \hat t/\hat R$, for several
values of the flow rate:  ($\blacksquare$) $\hat {Q}=0.063$~ml/s,
({\textcolor{red}{$\bullet$}}) $\hat {Q}=0.145$~ml/s,
({\textcolor{blue}{$\blacktriangle$}}) $\hat {Q}=0.19$~ml/s,
({\textcolor{green}{$\blacktriangleleft$}}) $\hat {Q}=0.3$~ml/s.
The experiments belong to Control sequence~$1$ (see table
\ref{experiments}). (b) Normalized width of the tip versus the
normalized displacement distance, $\hat U_0 \hat t/\hat R$, for
several values of the flow rate:
({\textcolor{green}{$\blacktriangleleft$}}) $\hat {Q}=0.13$~ml/s,
($\blacksquare$) $\hat {Q}=0.18$~ml/s,
({\textcolor{brown}{$\blacktriangledown$}}) $\hat {Q}=0.2$~ml/s,
({\textcolor{red}{$\bullet$}}) $\hat {Q}=0.31$~ml/s,
({\textcolor{blue}{$\blacktriangleright$}}) $\hat {Q}=0.47$~ml/s.
The experiments belong to the Reactive sequence (see
Table~\ref{experiments}).} \label{f.9}
\end{figure}

Fig.~\ref{f.9}a shows the normalized width of the finger, $W/\hat
d$, for the experiments in control sequence~$1$, where two
Newtonian fluids are displaced. For each flow rate, $W/\hat d$
saturates at values smaller than unity, indicating that in all
cases fluid~$2$ was only partially displaced. Qualitatively
similar results are given in \cite{Peti96}. Fig.~\ref{f.9}b shows
analogous results for the reactive displacements at different flow
rates. The interfacial instability clearly results in an efficient
mass transport in the cross flow direction. For each flow rate,
$W/\hat d$ saturates at values very close to unity. Observe that
the initial points on the time series are noisy, followed by a
rapid increase to near saturation, then slow approach to unity as
the wall layers are consumed by the secondary flows/instability.
The initial noisy part of the curves corresponds to
destabilization of the initial interfacial spike.

In comparing these two figures, recall that the time axis has been shifted to correspond to the initial appearance of fluid~$1$.
 In the case of the unstable displacements, this is harder to determine precisely due to the destabilized wispy spike.
  Once the bulk of the displacing finger arrives, the rapid increase to near the saturation values is similar in both cases.
   The wider spread of the curves in Fig.~\ref{f.9}b is due to the early arrival of the initial spike and slower erosion of the final wall layer.
   In terms of a volumetric measure of efficiency, the reactive displacement leaves $\lesssim 5\%$ of the displaced fluid volume,
   and this appears to be insensitive to the flow rate.
   The stable Newtonian displacements leave between $20\%$ and $60\%$ of the displaced fluid in the pipe,
   with strong dependence on the flow rate.

The key observations of Fig.~\ref{f.9}b is that all the efficiency curves (time shifted for arrival)
have similar shape and that the time scales with  $\hat{R}/\hat{U}_0$.
In terms of an averaged volumetric concentration of displaced fluid, say $\bar{C}(\hat{x},\hat{t}) = (W/\hat d)^2$, this suggests a functional form:
\[ \bar{C}(\hat{x},\hat{t}) = \bar{C}(\hat{x} -\hat{U}_0 \hat{t},\hat{t}\hat{U}_0/\hat{R}), \]
and the sigmoid shape is reminiscent of axial dispersion in a
moving frame of reference. If this dispersion is governed by an
axial diffusivity $\hat{D}_a$, the timescale for the spreading is
$\hat{t} \sim \hat{R}^2/\hat{D}_a$. Thus, the collapse of our data
with respect to the time variable $\hat{R}/\hat{U}_0$, suggests
the scaling law
\[ \hat{D}_a \sim \hat{R} \hat{U}_0  , \]
which infers a similar scaling law for a transverse diffusivity of mixing, $\hat{D}_r$, see \S \ref{sec:industrial}.

Even if redrawn volumetrically, the curves in Fig.~\ref{f.9}b are
not symmetric about the mid-concentration. Additionally, close
inspection reveals small  differences between the curves for
different flow rates. These suggest that, although axial
dispersion may provide a reasonable leading order model, the
dispersion coefficient will vary nonlinearly with respect to
$\bar{C}$ and have other weak dependencies. This is in fact
physically obvious: the mechanism governing the initial
instability of the wispy spike ($\bar{C} \sim 0$) is clearly quite
different to that governing erosion/break up of the residual wall
layer of C-940 ($\bar{C} \sim 1$).

\section{Hydrodynamic instabilities driven by viscosity gradients}
\label{sec:toymodel}

We would like to understand better the origin of the instabilities
that develop into secondary flows such as those in Fig.~\ref{f.6}.
Development of a complete stability analysis of the
reaction-diffusion system coupled to the hydrodynamics is a
formidable task, and perhaps unnecessary. One obstacle is that the
reaction kinetics are not fully quantified. Another obstacle is to
fully understand the coupling between the reaction and the
hydrodynamics. This analysis is however underway and we hope to
report the results later. For now, we offer some insights via
analysis of a simpler toy problem.

First, we note that the chemical reaction occurs on a time scale
$\hat t_{ch}$ which is much shorter than the hydrodynamic time
scales of the problem. This is supported by the estimation below.

Following ref. \cite{Atkins}, the characteristic time scale at
which the chemical reaction occurs is roughly $\hat t _{ch}
\approx 10^{-9}$s. In our experiments, the characteristic
hydrodynamic times are the diffusion time, $\hat t_D=\hat R^2/\hat
D $, the advection time, $\hat t_a=\hat R/\hat U_0$, and the
viscous time, $\hat t_v=\hat{\rho}\hat{R}^2/\hat{\eta}_0$. If one
considers $\hat D \approx 4.2 \cdot 10^{-12}~\text{m}^2/\text{s}$
(which is roughly the diffusion coefficient of fluorescein in our
solutions) and $\hat U_0 \approx 5~\text{mm/s}$ (which corresponds
to half of our experimental range), one can estimate $\hat t_D
\approx 7.2 \cdot 10^6$~s, $\hat t_a \approx 1.1$~s and $\hat t_v
\approx 0.3$~s. The numerical estimates above show a clear
separation of time scales in our problem: $\hat t_{ch}\ll \hat t_v
< \hat t_a \ll \hat t_D$. In view of this, it is reasonable to
assume that the fast reaction instantaneously establishes an
initial concentration profile when two fluids come into contact.
This "reaction front" is a diffuse layer of reacted fluid
separating the two bulk fluids, within which the $pH$ is
approximately neutral and the viscosity is consequently elevated.
In the following we will investigate the hydrodynamic stability of
such a viscosity profile.

This process is on the slow scale of $t_a$. Note that the reaction
is not auto-catalytic. Unless there is convective motion that
brings fresh un-reacted fluids into contact, the reaction front
broadens diffusively. No self-sustaining traveling chemical waves
arise. In addition, unlike the many "chemical reaction plus
buoyancy" studied referenced in \S \ref{sec:intro}, our fluids
have matching densities that are unchanged by the reaction.
Consequently, there is no momentum source resulting from the
reaction. Therefore, the direct effect of the continual reaction
is to create a source term in the advection-diffusion equation for
the concentration. Indirectly this modifies the viscosity profile.
For simplicity, however, we will ignore the effects of continual
reaction, and focus on the fate of the initial, chemically
created, concentration and viscosity profiles.
\begin{figure}[h,t]
\includegraphics[width=\textwidth]{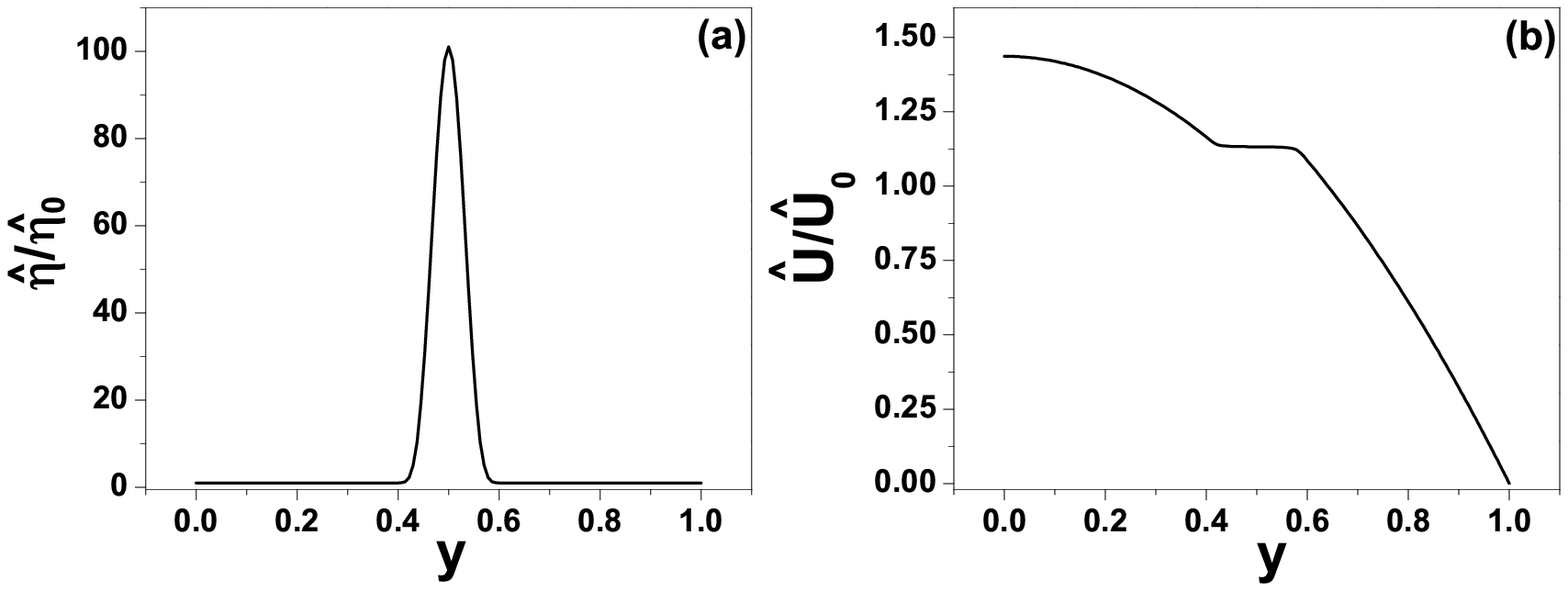}
\caption{Example basic viscosity and velocity profiles for $c_0=0.5$, $\Delta c_0 = 0.1$ and $a = 100$: (a) Normalized base
viscosity profile. (b) Normalized base velocity profile. }
\label{f.10}
\end{figure}

Furthermore we ignore the pipe geometry \footnote{Although the
onset of linear shear instabilities is different in pipe and plane
channel geometries, onset of interfacial instability is similar,
and the plane channel is simpler to deal with mathematically. Here
we study low $\Rey$ flows for which the non-interfacial modes are
assumed stable.}, and consider instead a symmetric plane
Poiseuille flow in a channel of width $2\hat{R}$ along which fluid
is pumped at mean speed~$\hat{U}_0$. The fluid is assumed to have
a concentration dependent viscosity $\hat{\eta}(c)$ that consists
 of a base viscosity $\hat{\eta}_0$ that is augmented (by a factor $a$)  over some finite range of concentrations,
  $\Delta c_0$, close to a fixed concentration value $c=c_0$. For example, one such function would be:
\[ \hat{\eta}(c) = \hat{\eta}_0 \left[ 1+ a  \cos^4 \left( \frac{\pi (c-c_0)}{2\Delta c_0}\right) \right], \]
which is depicted in Fig.~\ref{f.10}a. Ignoring the reaction, the
concentration satisfies an advection-diffusion equation and the
fluid flow satisfies the Navier-Stokes equations. There exists a
steady base flow in which the concentration varies linearly across
the channel. The increased viscosity tends to flatten the base
Poiseuille velocity profile within the diffuse layer, see
e.g.~Fig.~\ref{f.10}b.

We consider the linear stability of this base flow, using classical methods.
The stability problem is governed by the Reynolds number $\Rey = (\hat{\rho}\hat{R}\hat{U}_0)/\hat{\eta}_0 \lesssim 1$, the Schmidt number, $\Sc = \hat{\eta}_0/(\hat{\rho} \hat{D}) \in [10^4,10^6]$,
the initial concentration profile $C_0$,
the layer thickness and position, $\Delta c_0$ \& $c_0$,
and the amplitude $a$ of the viscosity jump.
Introducing the non-dimensional variables
$$x=\frac{\hat x}{\hat R},\quad y=\frac{\hat y}{\hat R},\quad t=\frac{\hat t}{\hat t_v},\quad U=\frac{\hat U}{\hat U_0}, \quad
\eta=\frac{\hat\eta}{\hat\eta_0}$$
 the growth or decay of a
linear mode $(f(y),c(y)) e^{i\alpha x+\sigma t}$ is governed by
the following eigenvalue problem: \bqa
\sigma\left( D^2-\alpha^2\right)f & = & \frac{1}{\Rey} L_1 f + L_2 f + L_3 c \label{oseq} \\
\sigma c & = & i\alpha\Rey\; \Bigl( (DC_0)f-U c\Bigr) + \frac{1}{\Sc}\left(D^2-\alpha^2\right)c \label{conceq}
\eqa
where $D=d/dy$, $\eta=\eta(C_0)$, $C_0 = 1-y$, and
\bqas
L_1&:=&-i\alpha\Bigl(U (D^2 - \alpha^2) -(D^2U)  \Bigr)\\
L_2&:=&(D^2\eta)\left(D^2+\alpha^2 \right) +2(D\eta)D\left(D^2-\alpha^2 \right) + \eta\left(D^2-\alpha^2\right)^2\\
L_3&:=& D^2\left(\frac{\pt\eta}{\pt C_0} DU\right) I+
  2D\left(\frac{\pt\eta}{\pt C_0} DU\right) D
 +\left(\frac{\pt\eta}{\pt C_0} DU\right) (D^2+\alpha^2)
\eqas We discretize the equations using Chebyschev polynomials and
determine the maximal growth rates $\sigma_R$ as a function of
wave number~$\alpha$ (and the other dimensionless parameters), in
the standard way, see e.g.~\cite{Schmid}. Examples of the results
are shown below.

\subsection{Results}

\begin{figure}
\unitlength 1cm
\includegraphics[width=\textwidth]{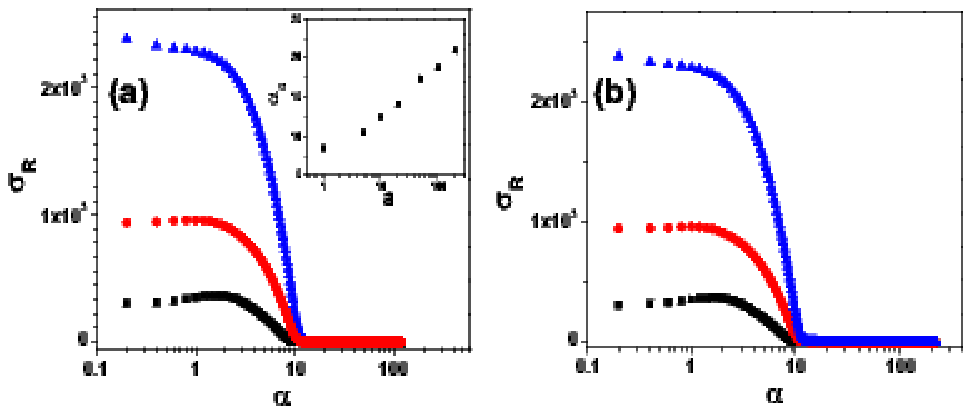}
\caption{Example results: $\sigma_R$ vs $\alpha$ for the base flow
of Fig.~\ref{f.10}  with the viscosity amplitudes
({\textcolor{blue}{$\blacktriangle$}})~$a=100$
($\blacksquare$)~$a=5$ and ({\textcolor{red}{$\bullet$}})~$a=20$:
(a) $\Sc = 10^4$. In the inset we display the dependence of the
critical wave number, $\alpha_c$, on the amplitude $a$; (b) $\Sc =
10^6$.} \label{f.11}
\end{figure}

Typically in our experiments we have $\Rey \lesssim 1$, so that we
are far below the range for inertia-driven shear instabilities. If
$a=0$, the viscosity is constant so that equation (\ref{oseq})
decouples and reduces to the classical Orr-Sommerfeld equation,
which is stable at low \Rey. Equation (\ref{conceq}) then gives
only stable modes with $\sigma_R < 0$ decreasing quadratically
with $\alpha$. As the amplitude $a$ increases, unstable modes are
found over an increasing range of $\alpha$, including the long
wave limit $\alpha = 0$, with diffusion stabilizing the short
wavelengths. Figure~\ref{f.11} shows typical examples of the
growth rate $\sigma_R$ vs $\alpha$, for a range of different $a$,
at two different values of \Sc.

The long wave limit $\alpha \to 0$ tends to give the largest
growth rates, but these modes will not be excited in a developing
finger-like displacement. Unstable wave numbers are found up to a
critical wave number $\alpha_c$, which varies with~$a$ and $\Delta
c_0$, increasing mildly with~$a$. A possible physical
interpretation of the ($x$-independent) long wavelength
instability is simply that a different base flow may exist and be
somehow more stable. In order to compare with the experimental
results, note that the wave numbers are scaled with $\hat{R}^{-1}$
and the growth rates with the inverse viscous timescale,
$\hat{t}_v$, (which is $\approx 0.3$~s for our experiments). Thus,
we can see from Fig.~\ref{f.11} that wavelengths $\sim \hat{R}$
are certainly excited at viscosity amplifications $a = 100$. The
growth rates are very rapid and we would therefore not expect to
see linear modal effects.

A complete exploration of the parameter space of this problem does not seem worthwhile and is not easy numerically. Although we typically have large $\Sc$, the limit $\Sc \to \infty$ results in loss of the diffusive terms in the concentration equation and it is these that stabilize the short wavelengths. The other limits of large $a$ and small $\Delta c_0$ will obviously eventually cause problems for the Chebyschev expansion, and would be better investigated analytically, as would the large and small wave number limits.

\section{Discussion and Conclusions}

In this study we have presented experimental evidence of an
inertial free shear flow instability caused by changes in the
local fluid rheology. We have focused on low $\Rey$ number
displacement flows in an horizontal pipe. An acid-base type
chemical reaction occurs near the interface between fluids and
results in molecular reorganization of the Carbopol-940 (C-940)
polymer. The main result of this molecular reorganization is the
formation of a stiff gel, characterized by a large viscosity and a
significant yield stress.

In order to test the effect of these rheological changes on the
hydrodynamic stability of the flow, we have carried out two
control experimental sequences, see Table \ref{experiments} series
$1$ and $2$. In neither of the control experiments was a flow
instability observed. If both fluids were Newtonian (first control
sequence), the displacement flow was dominated by a long finger of
fluid~$1$ penetrating into fluid~$2$, Fig.~\ref{f.4}. The flow was
stable and the second fluid was never completely removed from the
channel walls, Fig. \ref{f.9}a. If the displaced fluid was a yield
stress fluid, the flow remained stable at all times, (second
control sequence), see Fig. \ref{f.5}.

The behavior was significantly different in the reactive case. The
interface between the fluids becomes unstable and mixes across the
entire channel. The flow fields are unsteady and characterized by
a strong secondary motion in a direction orthogonal to the mean
flow direction, Fig.~\ref{f.7}b. The relevance of our experimental
findings is threefold.
\begin{enumerate}
  \item First, the coupling between chemical reaction and a strong \underline{non-monotonic}
  local change in the fluid rheology is novel. For simple fluids, viscosity variations can be
  caused by temperature or concentration gradients, but these tend to be gradual and monotone.
   Use of complex fluids allows one to localize the change in viscosity and produce non-monotone
   effects.\\
The instability mechanism we have described in this paper is not
restricted to interfacial flows, where acid-base type chemical
reactions take place at the interface between two fluids via
transport of unbalanced charges across the interface. Similar
chemical reactions, which result in significant changes in the
fluid rheology, may be triggered in a non invasive way, e.g. by
exposing fluid parcels to either electromagnetic radiation or heat
waves which will locally initiate a polymerization reaction. For
example, there exist numerous polymeric materials (epoxies) widely
used in photo lithography, whose viscosity increases dramatically
(up to solidification, depending on the chemical nature of the
polymer and the exposure conditions) upon exposure to
monochromatic electromagnetic radiation. Also, liquid silicone
elastomers may display a similar behavior (they turn from viscous
liquid state to solid elastic) when locally heated. This
observation, together with an instability mechanism similar to the
one illustrated in this paper, may open new possibilities of
externally controlling the hydrodynamic stability of inertia free
shear flows.
   %Here this was accomplished with a chemical reaction, but this is not the only method to accomplish this.
   % {\bf examples of the UV method, other e.g. fibres, or should we not give away any ideas here?}
  \item Secondly, we have demonstrated a significant increase in the displacement efficiency for reactive displacements.
  This has an impact on process applications such as these described in \S \ref{sec:industrial},
  and this was the initial motivation for our study.
  \item Thirdly, we have demonstrated a new mechanism for low $\Rey$ mixing that does not
  require large inertial energies.
  This in itself opens up many interesting avenues for
  practical applications, e.g. mixing in micro-fluidic flows.
Indeed, efficient mixing in the absence of inertia requires an
additional mechanism than molecular diffusion (which is the least
efficient). Several micro mixing techniques have been proposed
during the past half decade, all based on different mechanisms of
generating secondary flows (steady or random in time). One of
these techniques is based on the recently discovered elastic
turbulence ~\cite{etnature}, which is a random (in time) and
complex dynamic flow state in dilute solutions of linear flexible
polymers. As shown in~\cite{mixingnature, mixingPRE}, elastic
turbulence can be successfully employed in efficiently mixing
viscous fluids in the absence of inertia.
 Although the flow configuration we used in this study is not a typical mixing
configuration, based on the data presented in Figs.~\ref{f.6},
 and~\ref{f.8}, one can suggest that the instability discussed in this
paper could be alternatively used as a low $\Rey$ mixing method.

\end{enumerate}

Our hydrodynamic stability study is minimalist and we have
presented only brief results. However, the simple toy model shows
that long wavelength instabilities do result from non-monotonic
viscosity variations of the magnitude that we have and in the
correct range of $\Rey$ and $\Sc$. A more detailed study of these
reactive displacements is underway.

\begin{acknowledgments}
Financial support for this research was provided by the Natural Sciences and Engineering Research Council of Canada through strategic
project grant $306682$. This project is also supported by Schlumberger Oilfield Services and Trican Well Service Ltd.
We are grateful to Rastislav Seffer ($CBVL$, Vancouver) for valuable help and advice on the customization of stepping motor
controllers and software.
\end{acknowledgments}

\bibliographystyle{PoF}

\bibliography{instab}

\end{document}